\documentclass[fleqn,usenatbib]{mnras}

\usepackage{newtxtext,newtxmath}

\usepackage[T1]{fontenc}
\usepackage{ae,aecompl}

\usepackage{graphicx}	
\usepackage{amsmath}	
\usepackage{amssymb}	
\usepackage{comment}
\usepackage{color,soul}

\newcommand{\vir}[1]{``#1''}
\newcommand{\te}[1]{\text{#1}}

\newcommand{\su}{_\odot}
\newcommand{\gaia}{\textit{Gaia} }

\usepackage{etoolbox}
\makeatletter
\patchcmd\@combinedblfloats{\box\@outputbox}{\unvbox\@outputbox}{}{\errmessage{\noexpand patch failed}}
\makeatother

\title[The Galactic halo shape with \gaia DR2 RR Lyrae]{The
  shape of the Galactic halo with \gaia DR2 RR Lyrae. Anatomy of an
  ancient major merger}

\author[G. Iorio and V. Belokurov]{
Giuliano Iorio$^{1}$\thanks{giorio@ast.cam.ac.uk} and
Vasily Belokurov$^{1,2}$\thanks{vasily@ast.cam.ac.uk}
\\
$^{1}$Institute of Astronomy, University of Cambridge, Madingley Road, Cambridge CB3 0HA, UK\\
$^{2}$Centre for Computational Astrophysics, Flatiron Institute, 162 5th Avenue, New York, NY 10010, USA\\
}
\date{Accepted XXX. Received YYY; in original form ZZZ}

\pubyear{2018}

\begin{document}
\label{firstpage}
\pagerange{\pageref{firstpage}--\pageref{lastpage}}
\maketitle

\begin{abstract}

We use the \gaia DR2 RR Lyrae sample to gain an uninterrupted view of
the Galactic stellar halo. We dissect the available volume in slices
parallel to the Milky Way's disc to show that within $\sim30$ kpc from
the Galactic centre the halo is triaxial, with the longest axis
misaligned by $\sim70^{\circ}$ with respect to the Galactic
$x$-axis. This anatomical procedure exposes two large diffuse
over-densities aligned with the semi-major axis of the halo: the
Hercules-Aquila Cloud and the Virgo Over-density. We reveal the
kinematics of the entire inner halo by mapping out the amplitudes and
directions of the RR Lyrae proper motions. These are then compared to
simple models with different anisotropies to demonstrate that the
inner halo is dominated by stars on highly eccentric orbits. We
interpret the shape of the density and the kinematics of the \gaia DR2
RR Lyrae as evidence in favour of a scenario in which the bulk of the
halo was deposited in a single massive merger event.

\end{abstract}




\section{Introduction}

In $\Lambda$CDM Cosmology, Dark Matter halos are rarely spherical,
their shapes controlled by the environment and the accretion history
\citep[][]{Frenk1988,Dubinski1991,Warren1992,Colberg1999,Allgood2006,Bett2007,Hahn2007}. At
the early stages of the halo assembly, the shape is typically prolate
and aligned with the narrow filaments, via which the mass is supplied,
but with passing of time, halos can become triaxial or even oblate, as
the feeding filaments swell and the direction of accretion changes
\citep[e.g.][]{Cole1996,Tormen1997,Altay2006,Vera-Ciro2011,Libeskind2013}. This
metamorphosis does not necessarily imply that the memory of the early
halo configuration is completely erased. Instead, at redshift zero,
the history of the Dark Matter halo evolution may be deciphered by
studying how its shape changes with Galactocentric radius
\citep[e.g.][]{Hayashi2007,Vera-Ciro2011}. Note however, that
inclusion of baryons \citep[see
  e.g.][]{Kaza2004,Gnedin2004,Debattista2008,Abadi2010} or adaptation
of a different Dark Matter model
\citep[e.g.][]{Avila2001,Dave2001,Mayer2002,Peter2013} can affect the
details of some of the above calculations.

While the behavior of Dark Matter halos shows several coherent trends,
stellar halos appear to display a wider diversity, linked to the
strong suppression of star formation in low-mass Dark Matter clumps
\citep[see e.g.][]{Bullock2005, Cooper2010}. One of the important
corollaries of the above stochasticity is the expectation that the
bulk of the (accreted) stellar halo of a Milky Way-like galaxy is
contributed by a small number of massive dwarf galaxies \citep[see
  e.g.][]{DeLucia2008,Deason2013}. The picture therefore emerges in
which the most massive halo progenitors not only can define the shape
of the stellar halo \citep[see][]{Deason2013} but also set its overall
metallicity \citep[][]{Deason2016,DSouza2018}. It is difficult to
produce stellar halos without invoking (at least some of) the
processes that lead to formation of stars. The inclusion of baryonic
physics tends to alter the shapes of the resulting stellar halos
significantly. For examples, the inner portions of the stellar halos
built up with semi-analytic machinery are often prolate \citep[see
  e.g.][]{Cooper2010}, while hydro-dynamical simulations deliver
mostly oblate shapes \citep[e.g.][]{Monachesi2018}. The prevalence of
the oblate shapes in the simulated stellar halos is sometimes linked
to the significant contribution of so-called in-situ component
\citep[][]{Benson2004,Zolotov2009,McCarthy2012,Tissera2013,Cooper2015}.

Early attempts to gauge the properties of the Milky Way's stellar halo
had to rely on the small number of tracers and/or sparse sky coverage
\citep[][]{Preston1991,Reid1993,Sluis1998,Morrison2000,Siegel2002}. Recently,
thanks to the availability of wide-area deep imaging data, the shape
of the Milky Way's stellar halo has been the focus of many studies
\citep[e.g.][]{Newberg2006, Bell2008, JuricVOD,
  Sesar2011,deasonhalo,xuehalo,Iorio18}. While surveys like the SDSS
\citep[see][]{SDSSdr12} do provide a much broader view of the halo
\citep[see e.g.][]{Carollo2007, Carollo2010}, large swathes of the sky
are still missing, leaving portions of the inner Galaxy unmapped. Most
recently, the \gaia mission \citep[see][]{Gaia} has provided the first
all-sky view of the Galactic halo. By combining the variability
statistics from \gaia DR1 with the color information from \gaia and
2MASS, \citet{Iorio18} built a sample of $\sim$22,000 RR Lyrae
covering most of the celestial sphere, except for narrow regions close
to the Galactic plane. Using these old and metal-poor pulsating stars
and taking advantage of largely un-interrupted view of the Galaxy,
they were able to test a wide range of stellar halo models.

In this Paper, we aim to use the \gaia DR2 RR Lyrae stars to get
both closer to the centre of the Milky Way and to go further beyond
the reach of the \citet{Iorio18} analysis by linking the RR Lyrae density
evolution with their kinematics. Our study is motivated by the recent
discovery of tidal debris from what appears to be an ancient major
merger event \citep[][]{BelokurovSa,MySa,Helmi18}
Identified first in the Solar neighborhood, this debris cloud,
sometimes referred to as the ``Gaia Sausage'' has recently been shown
to dominate the Galactic stellar halo, stretching from regions in the
Milky Way's bulge to near and past the halo's break radius around
$20-30$ kpc \citep[see][]{Deason18,Simion18,Lancaster}.  
  Additional evidence has been found in the studies of the detailed
  chemical abundances of the nearby halo stars
  \citep[see][]{Hayes18,Haywood2018,Mack18}.  
  While many pieces of
the ``Gaia Sausage'' have already been reported in the literature,
here we attempt to provide the first comprehensive map of this {\it
  largest halo sub-structure}.  The kinematic portion of our study is
complementary to the work of \citet{Wegg18}, who recently used a
sample of PanSTARRS1 RR Lyrae stars to constrain the shape of the
inner portion of the Galactic gravitational potential.
 
This Paper is organized as follows. Section~\ref{sec:sample} describes
the construction of the clean sample of \gaia DR2 RR Lyrae stars.  In
Section~\ref{sec:slice} we show how these objects can be used to slice
the Galactic halo to reveal the remnant of a large dwarf galaxy buried
close to the Milky Way's centre. In Section~\ref{sec:disc} we
complement the spatial analysis with an all-sky RR Lyrae kinematic map and
we discuss the implications of our discovery. Finally, we summarise
the conclusions of this work in Section\ \ref{sec:conclusions}.

\section{The \gaia DR2 RR Lyrae sample}
\label{sec:sample}

\subsection{RR Lyrae stars in \gaia DR2}

The \gaia DR2 catalogue \citep{GaiaDR2,Gaia} contains 550,737 variable
stars, among which more than 200,000 are classified as RR Lyrae (RRL)
stars \citep{GaiaVariable}. In the \gaia data, the RRL stars are
spread across separate catalogues: some are in the tables reporting
the results of the general variability analysis
(e.g. \texttt{vari\_classifier\_result} and
\texttt{vari\_time\_series\_statistics}); others can be found in the
SOS (Specific Object Studies, \citealt{Clementini}) table
\texttt{vari\_rrlyrae}.  For further details on the variable stars
classification and analysis in \gaia DR2 please see
\cite{GaiaVariable}.  About half of the RRL stars are shared between
the SOS and the other tables, thus in order to merge all classified
RRL stars into one final sample, we joined the table
\texttt{vari\_classifier\_result} and the table \texttt{vari\_rrlyrae}
using the \textit{source\_id} unique identifier of each star.  The
final sample contains 228,853 RRL stars ($\sim77\%$ RRab, $\sim21 \%$
RRc and $~\sim2\%$ RRd).  Finally, this table containing all of the
Gaia DR2 RRL stars, was merged with the main \texttt{gaia\_source}
catalogue, thus complementing the RRL positions with photometry
and astrometry.

The top-left panel of Fig.\ \ref{fig:allsky} shows the distribution of
the RRL stars in our sample as a function of the Galactic coordinates
$(\ell, b)$.  A number of genuine as well as spurious objects stand
out against the underlying diffuse distribution of stars. These
include the Magellanic Clouds, the Sagittarius dwarf galaxy and its
tails, a number of other dwarf spheroidal satellites and globular
clusters, together with over-densities close to the Galactic plane. In
order to study the broad-brush properties of the stellar halo, we
first attempt to clean the sample and remove RRL stars belonging to
the most obvious compact structures, including both bona fide
satellites as well as data artefacts and contaminants. The following
section describes in detail the selection cuts applied to clean the
\gaia DR2 RRL sample.

\subsection{Cleaning}

\begin{figure*}
\centering
\centerline{\includegraphics[width=0.95\textwidth]{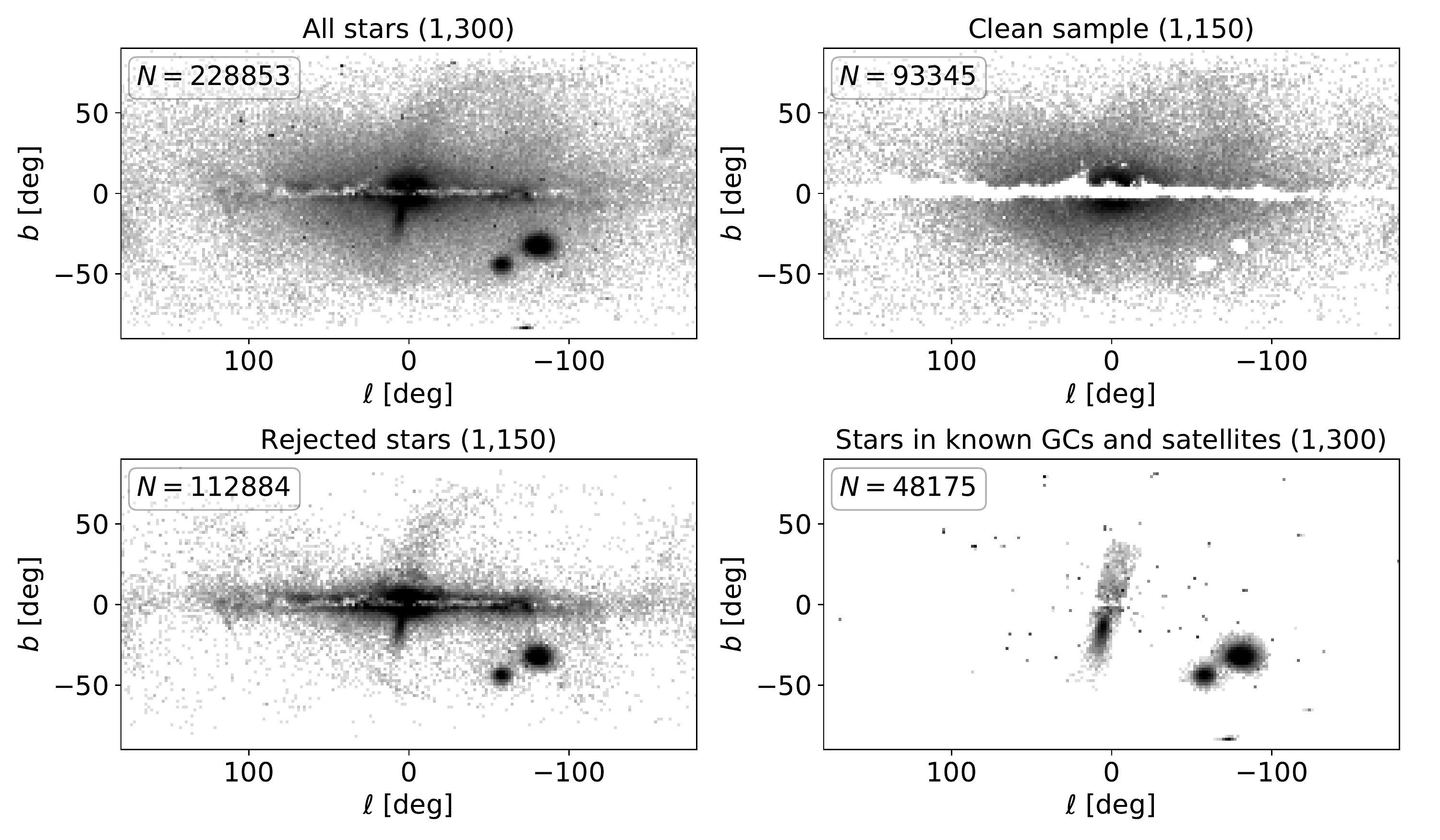}}
\caption[]{All-sky RR Lyrae density maps in Galactic coordinates
  ($\ell,b$).  \textbf{Top-left:} original sample of stars classified
  as RR Lyrae in \gaia DR2.  \textbf{Top-right:} clean sample of RR
  Lyrae stars used in this work.  \textbf{Bottom-left:} objects
  filtered from the original sample using a combination of photometric
  and astrometric selection cuts.  \textbf{Bottom-right:} stars in the
  original sample belonging to known globular clusters, dwarf
  spheroidal satellites, the Sagittarius dwarf and a part of its
  stream and the Magellanic Clouds.  The numbers in brackets in the
  title of each panel indicate the dynamic range of the grey-scale
  map; the number in the top-left corner indicates the total number of
  stars in each panel.}
\label{fig:allsky}
\end{figure*}

{\bf Artefacts and contaminants.} The percentage of contaminants in
the \gaia DR2 RRL sample has been estimated in \cite{GaiaVariable} and
in \cite{Clementini} using auxiliary catalogues. They found a low
level of contaminants ($5\%$-$15\%$) except in the region of the bulge
and in the area close to the Galactic disc where the contamination
level appears to increase up to many tens of per cent.  The
contaminants in these crowded fields are predominantly eclipsing
binaries and blended sources, with a tiny number of spurious
detections due to misclassified variable stars
\citep{GaiaVariable}. To cull the majority of the likely contaminants
we only keep objects that satisfy the following selection criteria:
\begin{itemize}
\item  $AEN<0.25$,
\item $BRE<1.5$,
\item $E(B-V)<0.8$.
\end{itemize}
The cut on the \textit{astrometric\_excess\_noise}, $AEN$, eliminates
unresolved stellar binaries, blends and galaxies \citep[see
  e.g.~][]{Koposov17} as their astrometric behavior deviates from that
of the single point sources \citep[see][]{Lind}.  The
\textit{phot\_bp\_rp\_excess\_factor}, $BRE$, represents the ratio
between the combined flux in the \gaia $BP$ and $RP$ bands and the
flux in the $G$ band, and thus by design is large for blended sources
\citep[see][]{EvansGaia}.  Finally, we remove stars in regions with
high reddening, $E(B-V)$ \citep[according to][]{dustext}, for which
the dust extinction correction is likely unreliable (see
Sec.\ \ref{sec:dust}).

{\bf Globular clusters and dwarf satellites.} We have removed all
stars within twice the half-light radius of all known globular
clusters listed in \cite{Globularcluster}.  Concerning the dwarf
spheroidal satellites, we excluded all stars within an angular
distance of $0.5^\circ$ from the centre of Carina, Ursa Minor and
Hercules, $0.3^\circ$ from Leo I and Leo II, and $1^\circ$ from
Sculptor, Fornax, Draco and Sextans.

{\bf Sagittarius dwarf.} First, we selected all stars with
$|\tilde{B}-\tilde{B}_\te{Sgr}|<10^\circ$ and $|\tilde{\Lambda} -
\tilde{\Lambda}_\te{Sgr}|<50^\circ$, where $\tilde{B}$ and
$\tilde{\Lambda}$ are the latitude and longitude in the coordinate
system aligned with the Sagittarius stream as defined in
\cite{BelokurovSgr} and $\tilde{B}_\te{Sgr}=4.24^\circ$ and
$\tilde{\Lambda}_\te{Sgr}=-1.55^\circ$ represent the position of the
Sagittarius dwarf. Then, among the selected objects, we got rid of all
stars with a proper motion relative to Sagittarius lower than $1.3
\ \te{mas} \ \te{yr}^{-1}$, using the dwarf's proper motion as
reported in \cite{HelmiGaia}. These spatial and kinematical selection
cuts appear to be effective in removing the Sagittarius dwarf and the
initial portions of its tails, as demonstrated in
Fig.\ \ref{fig:allsky}. We have decided against extending the filter
to the rest of the Sagittarius stream ($|\tilde{\Lambda} -
\tilde{\Lambda}_\te{Sgr}|>50^\circ$) to avoid over-cleaning our sample
at large Galactic latitudes.

{\bf Magellanic Clouds.} 
In order to identify (and remove) the stars that belong to the Clouds,
we selected all objects within an angular distance of $16^\circ$
($12^\circ$) from the LMC (SMC). Of these, only the stars with proper
motion relative to the LMC or to the SMC lower than $5 \ \te{mas}
\ \te{yr}^{-1}$ with respect to the values reported in \cite{LMCpm}
were retained for further consideration. Given the expected $G$
magnitude of the RRL stars at the distance of the LMC and SMC
\citep{Belokurov17}, we required the Magellanic stars to have
$18.5<G<20$ ($G$ corrected for the extinction, see
Sec.~\ref{sec:dust}).  Finally, in order to avoid the crowded central
part of the Clouds, we selected all stars within $5^\circ$ from the
LMC and SMC centre, independently of their proper motions and $G$
magnitudes.
The stars that survived all of the selection cuts above were removed
from our RRL sample.

The sky distribution of the likely artefacts discarded by the
astrometric and the photometric cuts described above is shown in the
bottom-left panel of Fig.\ \ref{fig:allsky}. As expected, most of the
contaminants are located close to the Galactic plane, while others can
be found in the crowded regions of the Magellanic Clouds and the
Sagittarius dwarf galaxy. The distribution of the (removed) stars
projected to lie within known Galactic satellites (globular clusters,
dwarf galaxies) is shown in the bottom right-panel. The stellar
density distribution in the final clean sample containing $\sim$93,000
RRL stars is shown in the top-right panel.

\section{Galactic halo shape with \gaia DR2 RR Lyrae}
\label{sec:slice}

\subsection{RR Lyrae distances}
\label{sec:dust}  \label{sec:dist}

In the analysis presented below, we use a left-handed Cartesian
reference frame $(x,y,z)$ centred in the Galactic centre as defined in
\cite{Iorio18}.  In this reference frame, the Sun is at $(8,0,0)
\ \te{kpc}$.
Despite the photometric variability, RRL stars have an almost constant
absolute magnitude in the visual \citep[see e.g.][]{Catelan2004}. Thus
having measured the apparent magnitudes $G$ (corrected for the dust
extinction), we can directly estimate the heliocentric distances as
\begin{equation}
\log\left(\frac{D\su}{\te{kpc}}\right)=\frac{G-M_G}{5}-2.
\label{eq:amag}
\end{equation}
Before applying Eq.\ \ref{eq:amag}, we correct the observed apparent
magnitudes for the dust extinction and estimate $M_G$ exploiting the
stars in our sample with robust parallax measurements.

{\bf Dust extinction.}  The $G$ band absorption $A_{G}=k_GE(B-V)$ for
the RRab stars has been estimated in the SOS \citep{Clementini}
through the empirical relation derived by \cite{piersimoni}.  We used
these stars to directly estimate the extinction coefficient
$k_{G}=A_{G}/E(B-V)$, where the reddening $E(B-V)$ is obtained from
the maps of \cite{dustext}. The derived $k_{G}$ distribution shows a
narrow peak at $1<k_{G}<3.4$.
We fit a Gaussian model to the measurements of stars inside this
$k_{G}$ range taking into account the errors with the help of the
extreme deconvolution technique \citep[ED,][]{extreme}. The best-fit
centroid is at $k_{G}=2.27$, i.e. a value slightly lower compared to
that reported previously \citep[e.g.][]{Belokurov17}. To summarise,
the observed apparent magnitudes of all stars in our sample have
been corrected using $A_{G}=2.27E(B-V)$.

\begin{figure}
\centering
\centerline{\includegraphics[width=\columnwidth]{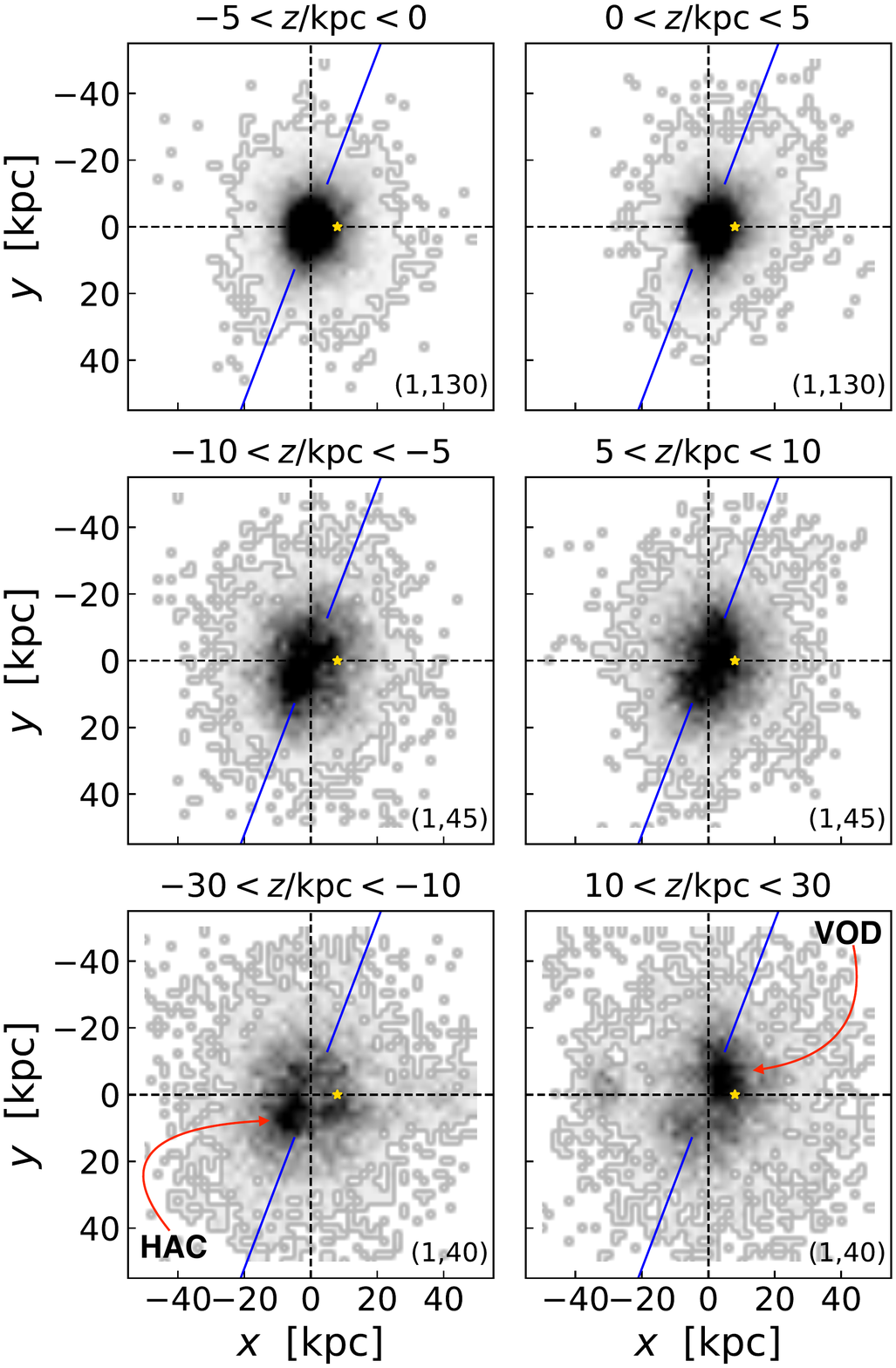}}
\caption[]{RR Lyrae density projected onto $x-y$ plane for three pairs
  of $z$ slabs. The two slabs in each pair are positioned
  symmetrically with respect to the Galactic plane.  The $z$ range is
  reported in the title of each panel. Yellow star indicates the
  position of the Sun ($x=8$ kpc, $y=z=0$ kpc). Dynamic range of the
  greyscale map is given in brackets in the bottom-right corner in
  each panel.  Blue lines indicate the orientation of the major axis
  of the triaxial stellar halo model described in
  \protect\cite{Iorio18}. The red arrows point out the location of the
  Hercules-Aquila Cloud (HAC) and the Virgo Over-Density (VOD).}
\label{fig:Zslice}
\end{figure}

{\bf Absolute magnitude in the $G$ band.} Inverting Eq.\ \ref{eq:amag}
and using the parallax $\varpi$, the absolute magnitude can be
estimated as
\begin{equation}
M_{G}=G  + 5\log \varpi    -10 .
\label{eq:dist}
\end{equation}
We selected 837 stars from the clean sample i) with robust estimates
of the parallax ($\varpi/\delta \varpi>10$) and ii) located in regions
where the correction for dust extinction is almost negligible,
i.e. $E(B-V)<0.1$. Before using Eq.\ \ref{eq:dist}, we corrected the
parallaxes for the offset $\varpi=0.029$ reported in \cite{Lind18}
\citep[see also][]{Muraveva18}.  The estimated $M_{G}$ distribution
exhibits a narrow peak for $|M_G|<1.5$. As in the case of the
reddening coefficient described above, the distribution was modeled as
a Gaussian taking into account the errors in $\varpi$ using the ED
machinery. The best-fit centroid is at $M_G=0.64$ and the best-fit
dispersion is $\sigma_{M_G}=0.25$.
We estimated the possible systematic effects repeating the analysis
considering only the stars classified as RRab ($M_G=0.64$) or RRc
($M_G=0.60$) in the SOS and considering only the metal poor
$\te{[Fe/H]}<-1.0$ ($M_G=0.58$) and the metal rich $\te{[Fe/H]}>-1.0$
($M_G=0.72$) stars\footnote{We used the metallicities reported in the
  SOS catalogue, see \cite{Clementini} for further details}.  We also
repeated the analysis not correcting for the parallax offset
($M_G=0.45$) or using the distance estimated from the parallaxes in
\cite{Bailer18} ($M_G=0.66$). Overall all the systematic differences
are within the measured spread in $M_G$.
For the rest of the paper, we consider $M_G=0.64$ for all the stars in
our clean sample. Note that a difference in magnitude of 0.25
corresponds to an uncertainty in distance of $\sim$1 kpc for a star at
10 kpc and  $\sim$4 kpc for a star at 50 kpc. Using
apparent magnitudes corrected for dust extinction and applying the
assumed (constant) value of the absolute magnitude, we estimated the
distance (Eq.\ \ref{eq:amag}) and the Galactocentric Cartesian
$(x,y,z)$ coordinates of each RRL star in our clean sample.

\begin{figure*}
\centering
\centerline{\includegraphics[width=0.95\textwidth]{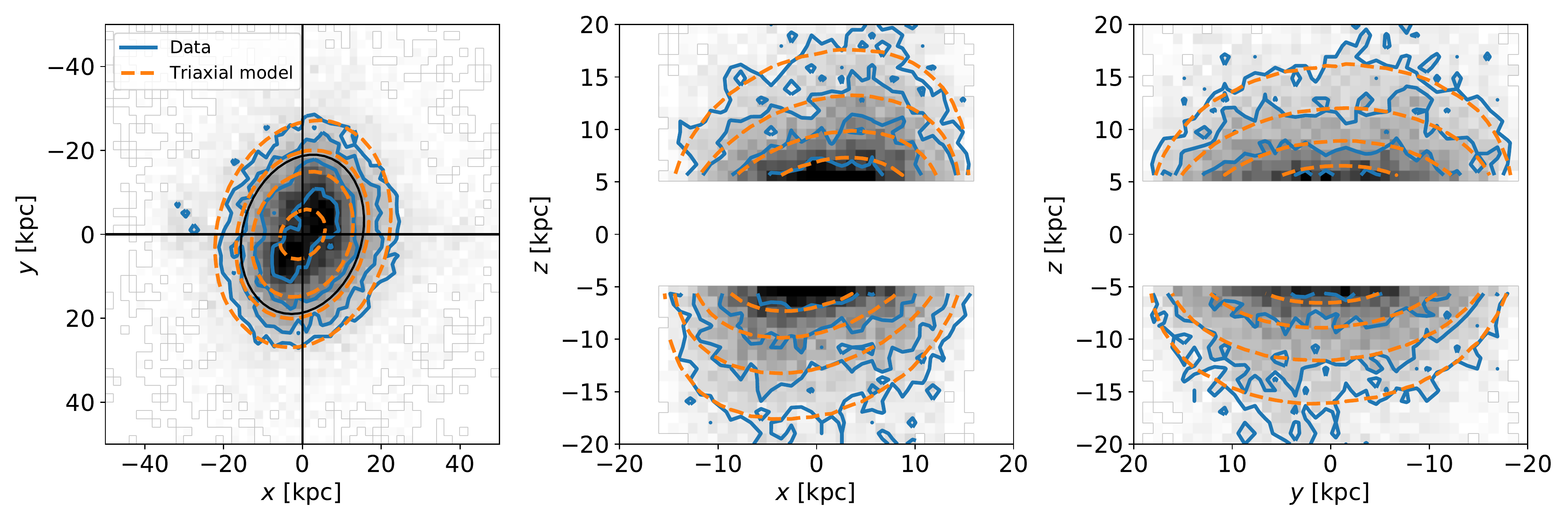}}
\caption[]{Three different projections of RR Lyrae density in the
  inner halo. Normalized star-counts ($N/N_\te{tot}$) are shown
  projected onto the principal planes in Galactic coordinates. Only
  stars with $5<|z/\te{kpc}|<30$ are included. The \gaia DR2 data (triaxial
  model) are shown as a greyscale map and blue contours (orange
  contours, see text for details). The contours are at $N/N_\te{tot}=
  (0.5, 1, 2, 4)\times10^{-4}$. {\bf Left-hand:} $x-y$ plane, the
  black ellipse has a major-to-minor axial ratio of $p=1.3$ and the
  major axis ($a=19.5\ \te{kpc}$) is tilted by $21^\circ$ with respect
  to the $y$-axis. {\bf Middle:} $x-z$ plane density of stars within
  the ellipse shown in the left-hand panel; {\bf Right-hand:} $y-z$
  plane density of stars within the ellipse shown in the left-hand
  panel.}
\label{fig:View}
\end{figure*}

\subsection{Slicing the halo} \label{sec:distribution}


Fig.\ \ref{fig:Zslice} shows the distribution of stars in the Galactic
$x-y$ plane in different $z$-slices.
Close to the Galactic disc ($0<|z/\te{kpc}|<5$), the distribution of
stars is symmetric with respect to the Galactic plane and most of the
stars can be found within 10 kpc from the Galactic centre.  At the
intermediate heights ($5<|z/\te{kpc}|<10$), the distribution preserves
the semblance of symmetry with respect to the Galactic plane, but the
bulk of the stars is now arranged within an elongated structure of
roughly ellipsoidal shape.  The major axis of this structure is tilted
by about $70^\circ$ (anti-clockwise) with respect to the Galactic
$x$-axis.  At even larger heights ($10<|z/\te{kpc}|<30$), two
over-dense regions are clearly visible: one above the Galactic plane
in the quadrant with positive $x$ and negative $y$ and one below the
Galactic plane, in the quadrant with negative $x$ and positive
$y$. The structure above the Galactic plane is the Virgo Over-Density
\citep[VOD,][]{VivasVOD, NewbergVOD, DuffauVOD, JuricVOD, BonacaVOD},
while the one below is the Hercules-Aquila Cloud
\citep[HAC,][]{BelokurovHAC,SimionHAC}. Note that these two striking
over-densities appear to lie at very similar distances from the
Galactic centre.

Apart from the VOD and the HAC, other minor substructures can be seen
in the last two $z$-slices: a circular overdensity of stars centred at
$x\sim-30$ kpc above the Galactic plane and a feeble overdensity strip
between $x\sim15$ kpc and $x\sim40$ kpc below the Galactic plane. Both
of these are the remaining portions of the Sagittarius stream at large
(in absolute values) Galactic latitudes (see Sec.\ \ref{sec:sample}).
We stress that the residuals of the Sagittarius stream cannot be
responsible of the elongated structure shown in
Fig.\ \ref{fig:Zslice}, in fact, the signature of the stream runs
almost exactly perpendicular to it, both above and below the Galactic
plane.  The possible random and systematic errors due to the
assumption of a constant $M_G$ act akin to spatial smoothing (of the
order of $1-4$ kpc) in the distribution of stars shown in
Fig.\ \ref{fig:Zslice}. This smearing effect should not have any
preferential direction, thus we expect that, in the ideal case of
perfect distance measurements, the elongated density structure ought
to emerge even more clearly.
In conclusion, we believe that the shape and the orientation of the
RRL distribution shown in Fig.\ \ref{fig:Zslice} represent faithfully
the intrinsic properties of the inner Galactic stellar halo.

The presence of a similar elongated density distribution of stars in
the halo was claimed previously by \cite{Newberg2006} and, most
recently by \cite{Iorio18} who used shallower ($G<17.1$) sample of RRL
star candidates from~ \gaia DR1. The best halo model reported in
\cite{Iorio18} is triaxial with a flattening along the direction
normal to the Galactic disc and the major axis lying in the Galactic
plane. The direction of the major axis is overplotted in the $x-y$
density maps in Fig.\ \ref{fig:Zslice}. Evidently, the ellipsoid's
orientation inferred by \cite{Iorio18} is compatible with the
elongation of the RRL distribution at intermediate heights and is
aligned with the direction connecting the VOD and the HAC structures.

Fig.~\ref{fig:View} presents a detailed comparison between the data
and the model similar to that introduced in \cite{Iorio18}. The
left-hand panel of the Figure shows the $x-y$ projection of the
distribution of the inner halo stars with $5<|z/$kpc$|<30$. The middle
(giving the $x-z$ view) and right-hand ($y-z$) panels only concern the
21,269 stars within the cylinder with $5<|z/$kpc$|<30$ and the base
shown as a black ellipse in the left panel of the Figure. The stellar
distributions in the last two planes are clearly different. In
particular, the iso-density contours in the $x-z$ plane are not
aligned with the Galactic axes.  It is interesting to explore whether
the distribution of stars whose projections on the three principal
planes shown in Fig.\ \ref{fig:View} can be reproduced by a smooth
triaxial halo model. For this purpose, we over-plot the iso-density
contours of the best-fit model labeled SPL-TR$^{qv}$ in
\cite{Iorio18}.  The original model is triaxial, with the major axis
lying in the Galactic plane and the minor axis pointing along the
Galactic z-axis. The ratio of major to intermediate axis is 1.27 and
the flattening along the vertical direction (i.e. minor axis)
decreases with the elliptical radius (i.e. the halo becomes more spherical). 
In order to describe better the
tilted distribution of stars in $x-z$ plane, the model was modified by
adding a rotation around the major axis of $\beta=20^\circ$ (see
\citealt{Iorio18} for a detailed description of the tilt angles
formalism).  Curiously, this model is similar to the best-fit triaxial
halo model found in \cite{Newberg2006}, except for the variation of
the vertical flattening.  The match between the data and the model is
impressive, especially considering that the fit was performed
exploiting only a limited portion of the halo's volume
\citep{Newberg2006,Iorio18}. Note, however, that according to the
left-hand panel in Fig.\ \ref{fig:View}, it is not possible to
reproduce the double peak in the projection of the density onto the
$x-y$ plane. These two off-centered over-densities correspond to the
VOD and the HAC.

Overall, the picture that emerges from Fig.\ \ref{fig:Zslice} and
Fig.\ \ref{fig:View} is that of a triaxial stellar halo,  misaligned
with respect to the $x$-axis and possibly misaligned with respect to
the Galactic plane. The latter statement is less certain due to the
presence of two large stellar over-densities, the VOD (in the Galactic
North) and the HAC (in the South). These large and diffuse Clouds
nonetheless appear to be connected to the stellar halo's ellipsoid,
being arranged approximately along its major axis.

\begin{figure*}
\centering
\centerline{\includegraphics[width=0.9\textwidth]{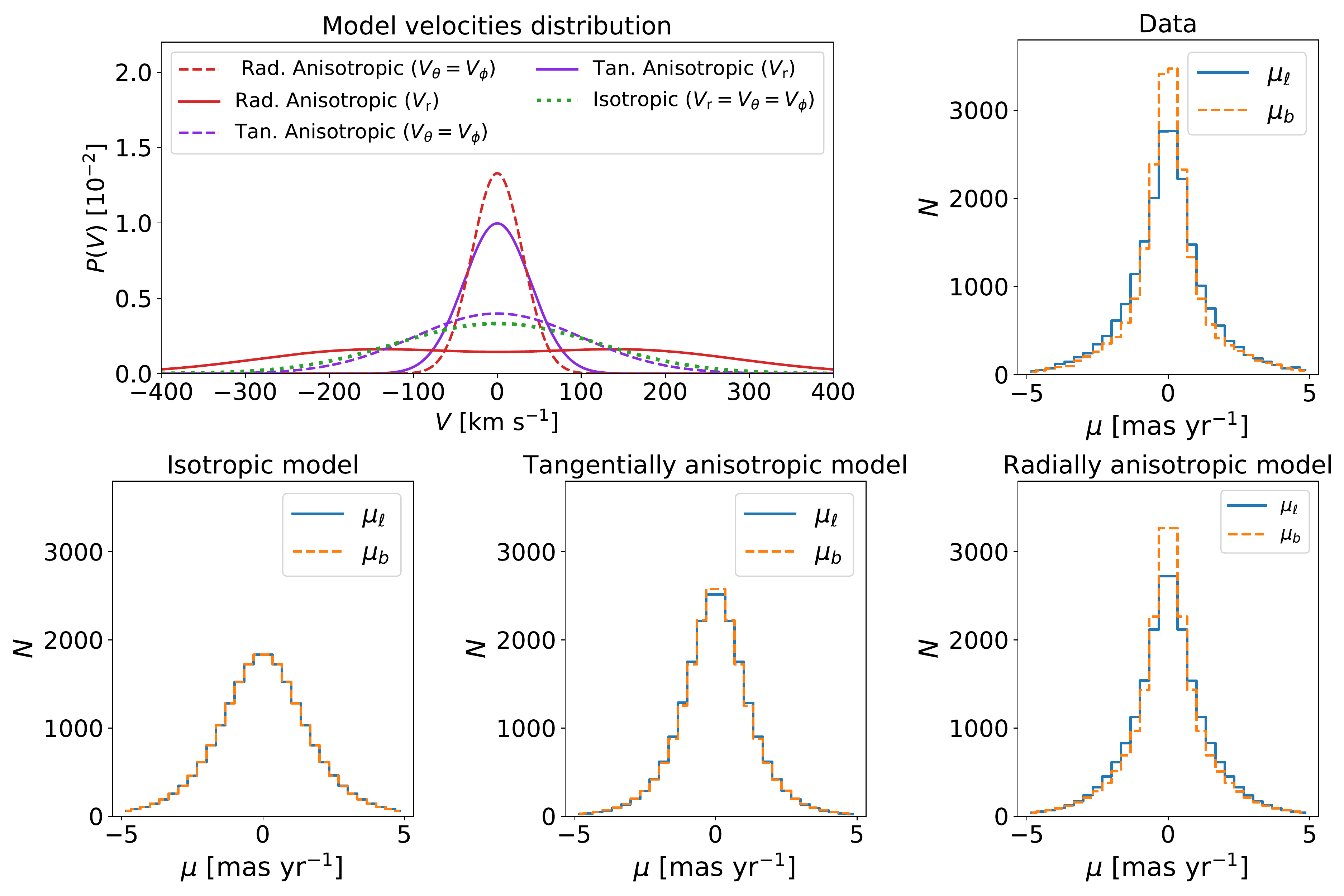}}
\caption[]{Galactic proper motion distributions in the \gaia DR2 RR
  Lyrae sample and 3 different halo models. Isotropic ({\bf
    bottom-left panel}), tangentially anisotropic ({\bf bottom-middle
    panel}) and radially anisotropic ({\bf bottom-right panel}) models
  are shown. These can be compared to the observed ($\mu_\ell$,
  $\mu_b$) distributions given in the {\bf top-right panel}. Only
  stars in the clean sample with $5<|z/\te{kpc}|<30$ and located
  within the ellipse shown in the left-hand panel of
  Fig.\ \ref{fig:View} are considered ($N$$\sim$21,000). The model
  proper motions are obtained by assigning to each star in our sample
  a Galactocentric spherical velocity $(V_\te{r}, V_\theta, V_\phi)$
  randomly drawn from the corresponding distributions shown in the
  {\bf top-left panel}.  The model histograms are acquired by taking
  the averaging deviation of 10,000 model realisations.
   The dispersion of this realisation set is negligible
    (less than
    $1\%$ at the histogram peaks) and it is not shown.
  All proper motions have been corrected for the Solar
  reflex motion \citep{rsunsch}.}
\label{fig:PM}
\end{figure*}

\section{Anatomy of an ancient major merger}
\label{sec:disc}

Confronted with the image of the stretched, likely triaxial inner
stellar halo as revealed in Figures ~\ref{fig:Zslice} and
~\ref{fig:View}, it is difficult to envisage a scenario in which such
a structure emerges naturally as a result of multiple randomly
oriented accretion events. On the other hand, major mergers tend to
produce exactly such a configuration of debris - an ellipsoid
elongated in the direction of the collision
\citep[e.g.][]{Moore2004,Cooper2010}. Even if the post-merger halo is
nearly axi-symmetric e.g. prolate, the shape can evolve with time,
under the influence of the growing baryonic component \citep[see
  e.g.][]{Debattista2008,Bryan2013,Tomassetti2016}.

Coincidentally, weighty evidence for an old accretion event of a
massive satellite on a low angular momentum orbit has recently been
uncovered in the Milky Way
\citep[see][]{BelokurovSa,MySa,Helmi18,Haywood2018}. 
The give-away sign that the object that merged with the Galaxy was massive is the metallicity
distribution of its tidal debris which reaches [Fe/H]$\sim-1$ (or
perhaps, even higher). The independent confirmation of the
progenitor's high mass comes from the discovery of a large number of
Globular Clusters likely associated with the event
\citep[e.g.][]{MyGC,Kraken}. The geometry of the impact can be gleaned
from the orbital anisotropy of the accreted stars: typical values of
$\beta\sim 0.9$ (the extreme stretching of the velocity ellipsoid
gained this halo component the name ``Gaia Sausage'') correspond to
eccentricities not too far from 1. This head-on collision appears to
have sprayed stars all over the Milky Way, dominating the stellar halo
out to $\sim30$ kpc as evidenced from the recent studies of the
orbital properties of a variety of halo tracers across a wide range of
Galactocentric distances
\citep[][]{Bird2018,Deason18,Lancaster}. Finally, using a sample of
RRL stars with radial velocities, \cite{Simion18} found that the HAC
and VOD can also be linked to this event, representing some of the
less mixed portions of its tidal debris. In the face of the evidence
above, the obvious question is whether some or most of the stars in
the triaxial structure revealed here (see
Sec.\ \ref{sec:distribution}) belong to the ``Gaia Sausage''.

With this conjecture in mind, we set out to explore the kinematics of
the RRL stars in the inner portions of the Milky Way's halo.

\begin{figure*}
\centering
\centerline{\includegraphics[width=0.95\textwidth]{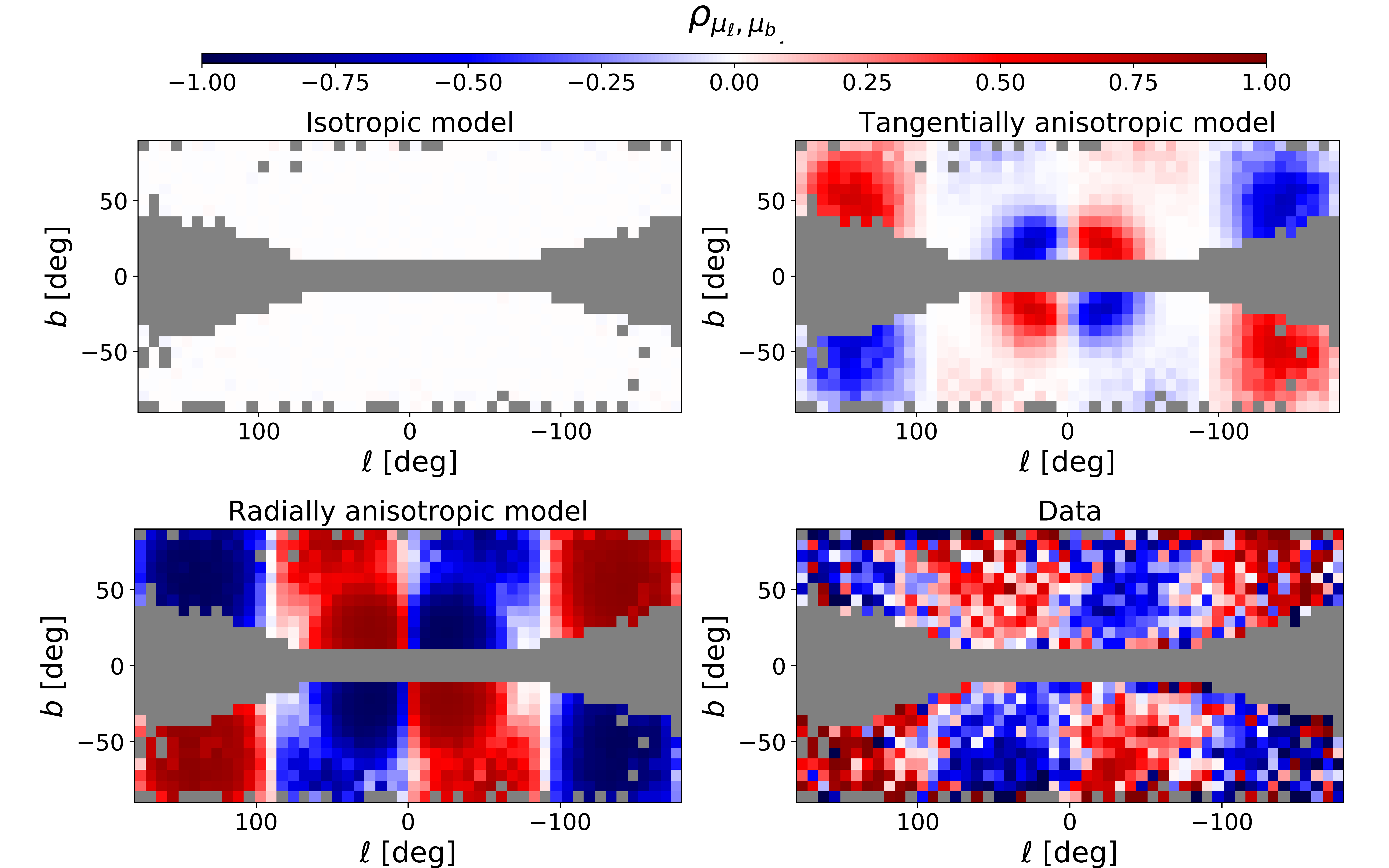}}
\caption[]{Pearson correlation coefficient of the Galactic proper
  motion components ($\mu_\ell, \mu_b$) as a function of the sky
  position in Galactic coordinates for the three models (isotropic,
  {\bf top-left panel}; tangentially anisotropic, {\bf top-right
    panel}; radially anisotropic, {\bf bottom-left panel} and the data
  {\bf bottom-right panel}). Only the stars in the clean sample with
  $5<|z/\te{kpc}|<30$ and located within the black ellipse shown in
  the left-hand panel of Fig.\ \protect\ref{fig:View}
  ($N$$\sim$21,000) are considered. The underlying model velocity
  distributions are shown in the top-left panel of
  Fig.\ \protect\ref{fig:PM}. The maps of the model correlation have
  been obtained by averaging the results of 10,000 model realisations.
The mean dispersion of the model realisations ($\sigma
  \rho_{\mu_\ell,\mu_\te{b}}\sim0.25$) is compatible with the level of
  random fluctuations displayed by the data.
  The proper motions have been
  corrected for the Solar reflex motion. The gray regions indicate
  bins without stars. }
\label{fig:PMmap}
\end{figure*}

\subsection{RR Lyrae kinematics} \label{sec:kine}
\label{sec:kinem}

Unfortunately, with only the proper motions measurements at our
disposal, we cannot directly study the orbital properties of the RRL
stars in our sample.  Recently \cite{Wegg18} circumvented this
limitation using a sample of PanSTARRS1 RRL stars in \gaia DR2 and
marginalising over the unknown line-of-sight velocity component. In
this work, we decided to use a simpler, but nonetheless effective and
illuminating approach.  In particular, we compare the observed proper
motions distributions to the predictions of several simplified halo
models.  More precisely, we generate mock proper motions values for
the stars in our sample assuming i) complete orbital isotropy, ii)
high radial anisotropy (similar to the ``Gaia Sausage'') as well as
iii) strong tangential anisotropy.  The model velocity distributions
in Galactocentric spherical coordinates are shown in the top-left
panel of Fig.\ \ref{fig:PM}. 
The isotropic model is composed of Gaussian velocity
  distributions with
  $\sigma_\theta=\sigma_\phi=\sigma_\te{r}=120\ \te{km}
  \ \te{s}^{-1}$.  The radially anisotropic model is inspired by that
  used in \citet{Necib18} and \citet{Lancaster}. Namely, the
  $V_\te{r}$ distribution contains two Gaussians with peaks at
  $V_\te{r}=\pm 160\ \te{km} \ \te{s}^{-1}$ and
  $\sigma_\te{r}=130\ \te{km} \ \te{s}^{-1}$, while the $V_\theta$ and
  $V_\phi$ distribution are single Gaussians centred in 0 and with
  $\sigma_\theta=\sigma_\phi=30\ \te{km} \ \te{s}^{-1}$. Therefore,
  this model corresponds to the extreme orbital anisotropy of
  $\beta\sim0.96$, similar to that observed for the ``Gaia Sausage''
  stars \citep{Belokurov17,MySa,Simion18}.  The tangentially biased
  model uses Gaussian velocity distributions with
  $\sigma_\theta=\sigma_\phi=100\ \te{km} \ \te{s}^{-1}$ and
  $\sigma_\te{r}=40\ \te{km} \ \te{s}^{-1}$, these values are
  arbitrary and have been set to reproduce a strong tangential
  anisotropy ($\beta=-5.25$) maintaining roughly the same velocity
  range of the isotropic and radially anisotropic models (see
  Fig.\ \ref{fig:PM}).  Note, however, that $\beta$ of our
  tangentially anisotropic model is somewhat lower (signifying a
  stronger anisotropy) than the lowest anisotropy estimates based on
  stellar tracers in the Milky Way stellar halo.
	\citep[see e.g.\ ][]{Kafle12,CKIII}.
In the analysis that follows, we focus only on the
$\sim$21,000 stars belonging to the triaxial structure highlighted in
Sec.\ \ref{sec:distribution}, i.e. those within the region with $5<|z/
\te{kpc}|<30$ and inside the black ellipse shown in the left-hand
panel of Fig.\ \ref{fig:View}. In practice, for each RRL star in our
sample, we draw $(V_\te{r}, V_\theta, V_\phi)$ randomly from the
distributions discussed above. Then, given the fully defined positions
and velocity vectors, we derive the model proper motions ($\mu_\ell,
\mu_b$). The final model predictions (see below) are obtained by
averaging over 10,000 model realisations.

Armed with the halo models, we compress the kinematic information into
two simple plots. First, as shown in Fig.\ \ref{fig:PM} we compare the
1-D distributions of $\mu_\ell$ and $\mu_b$ across the entire sample
of stars considered. From isotropic, through tangentially to radially
anisotropic models, the proper motion distributions become
narrower. The thinnest peak belongs to the distribution predicted by
the radially anisotropic model, which also possesses sizeable
wings. For the isotropic and the tangential case, the shapes of the
$\mu_\ell$ and the $\mu_b$ histograms are indistinguishable.  However,
for the radially biased model, the distributions of the two proper
motion components are clearly different: the $\mu_b$ histogram is more
peaked with respect to that of $\mu_\ell$. Most importantly, there is
a strikingly good match between the radially biased model and the data
shown in the top-right panel of Fig.\ \ref{fig:PM}.

Fig.\ \ref{fig:PMmap} extends the comparison from 1-D to 2-D,
i.e. demonstrates the behavior of the three models and the data as a
function of the position on the sky. Here we show the evolution of the
Pearson correlation coefficient
$\rho_{\mu_\ell,\mu_b}$\footnote{$\rho_{\mu_\ell,\mu_b}=\te{cov}(\mu_\ell,\mu_b)/\sigma_{\mu_\ell}\sigma_{\mu_b}$,
  where $\te{cov}(\mu_\ell,\mu_b)$ is the proper motions covariance
  and $\sigma_{\mu_\ell}$ and $\sigma_{\mu_b}$ are the proper motion
  standard deviations.} in bins of Galactic coordinates. The thus
calculated correlation can be interpreted as the characteristic
direction of motion on the plane of the sky.  No preferred direction
is expected for the isotropic model as confirmed in the top-left panel
of Fig.\ \ref{fig:PMmap}. In spherical polars, the velocity ellipsoids
of the tangentially and radially biased models are rotated by $\pi/2$
with respect to each other. As the top-right and the bottom-left
panels demonstrate, even in the absence of the third velocity
component, the stretching and the orientation of the velocity
ellipsoid can be deduced from its projection on the sky (for
reasonably nearby stars as those considered here). In the presence of
strong orbital anisotropy, the proper motions are heavily
correlated. The sign of the correlation depends on sky position and
changes every $90^\circ$ both in $\ell$ and $b$ creating the
characteristic \vir{red-blue} patterns visible in the top-right and
bottom-left panels of the Fig.\ \ref{fig:PMmap}. Without any doubt,
the observed pattern of the RRL motion shown in the bottom-right panel
matches that of the radially-anisotropic model (bottom-left).

We stress that the predictions of our simple kinematical models
  have converged, i.e. they do not change when the number of
  realisations is increased. In particular, the dispersion of the
  model representations is negligible in Fig.\ \ref{fig:PM} (less than
  $1\%$ at the histograms peaks) and the mean dispersion in the
  correlation maps in Fig.\ \ref{fig:PMmap}
  ($\sigma_{\rho_{\mu_\ell,\mu_\te{b}}}\sim0.25$) is compatible with
  the level of random signal oscillation shown in the data. In fact,
  even when using a single realisation the relevant features shown in
  Fig.\ \ref{fig:PM} and Fig.\ \ref{fig:PMmap} are still evident,
  although the overall proper motion distributions and correlation
  maps appear noisier. 

\begin{figure*}
\centering
\centerline{\includegraphics[width=0.9\textwidth]{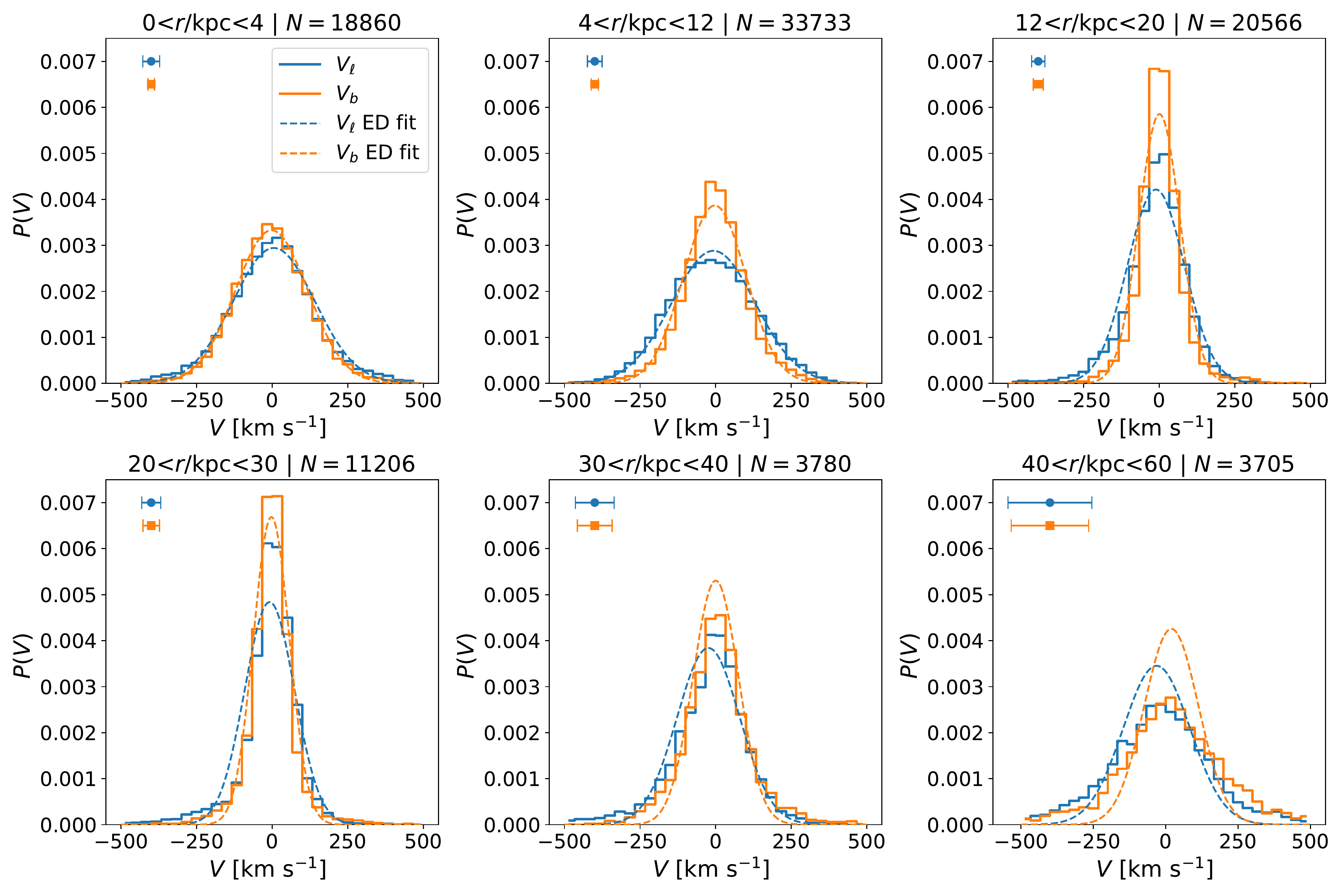}}
\caption[]{Distributions of the transverse velocities along the
  direction of the Galactic longitude ($V_\ell$, blue histograms) and
  latitude ($V_b$, orange histograms) for the RR Lyrae stars in the clean
  sample.  The panels show the velocity distributions in different
  radial bins as reported in the title of each panel together with the number of stars ($N$) in the bin. 
  The velocities have been corrected for the
  Solar reflex motion \citep{rsunsch}.
  The error bars in the top-left corners
  indicate the median velocity errors in each bin (blue-dot for
  $V_\ell$, orange-square for $V_b$). The velocity
  errors have been estimated by means of Monte Carlo simulations
  taking into account the observed proper motion errors and the
  estimated dispersion in the RR Lyrae absolute magnitude distribution (see
  Sec.\ \ref{sec:dist}). The dashed lines represent the
  results of Extreme Deconvolution Gaussian fits \citep{extreme}
  taking into account the errors on $V_\ell$ and $V_b$. 
   }
\label{fig:VlVbr}
\end{figure*}

\subsection{Radial trends}

\begin{figure*}
\centering
\centerline{\includegraphics[width=1.0\textwidth]{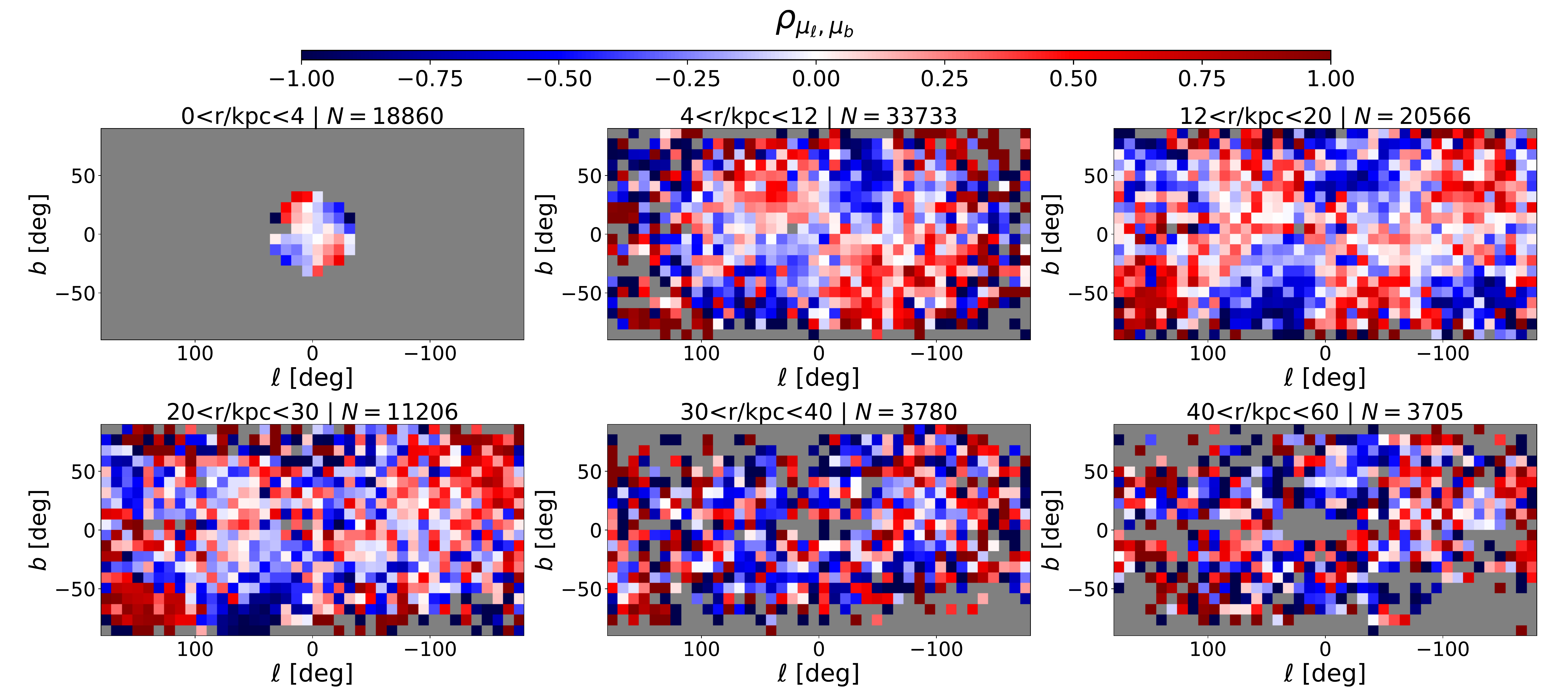}}
\caption[]{Same as Fig.\ \ref{fig:PMmap}, but considering all RR Lyrae
  stars in the clean sample in different radials bins as reported in
  the title of each panel  together with the number of stars in the bin. 
  The gray regions indicate bins without
  stars. Note that the strongest negative correlation (dark blue
  color) in the fourth radial bin (20-30 kpc) is supplied by the
  remaining portion of the Sgr trailing stream.}
\label{fig:Pcoffr}
\end{figure*}

This sub-section explores the dependence of the RRL kinematics on the
Galactocentric distance. In particular, we replicate the analysis
presented above using sub-samples of RRL  in different
Galactocentric distance $r$ bins.  Given the (maximal) spatial
blurring due to the uncertainties in the distance estimate
(Sec.\ \ref{sec:dist}), the minimal size of a distance bin is 4 kpc.
For each radial bin, we show both the 1-D distributions of the
transverse velocities derived from the~\gaia DR2 proper motions
(Fig.\ \ref{fig:VlVbr}) and the 2-D maps of the correlation
coefficient of the RRL proper motions on the sky
(Fig.\ \ref{fig:Pcoffr}).

{\bf Velocity distributions evolution.}  The amplitude of the proper
motion depends both on the velocity of a star as well as its distance
from the observer (Eq.\ \ref{eq:dist}). Therefore, we convert the
proper motions into physical velocities $V_\ell=K\mu_\ell D_\odot$ and
$V_b=K\mu_b D_\odot$ to track changes in kinematics across a wide range of
$r$. Here, $V_\ell$ and $V_b$ are the velocities along the direction
of the Galactic longitude and Galactic latitude and $K \approx
4.70$ is the conversion factor from $\te{mas} \ \te{kpc}
\ \te{yr}^{-1}$ to $\te{km} \ \te{s}^{-1}$.  The distributions of
$V_\ell$ and $V_b$ as a function of the Galactocentric distance are
given in Fig.\ \ref{fig:VlVbr}. In the bin closest to the Galactic
center, the two distributions are almost identical, in agreement with
the isotropic model (see the bottom-left panel in Fig.\ \ref{fig:PM}).
At the intermediate radii ($4<r/\te{kpc}<30$), the $V_\ell$ and $V_b$
distributions become progressively narrower; moreover, significant
differences between $V_b$ and $V_\ell$ histograms start to appear.
Thus, the kinematics of the RRL stars in this radial range appears
compatible with the expectation from the radially anisotropic model
(see the bottom-right panel in Fig.\ \ref{fig:PM}).  Beyond $30$ kpc,
the trend is reversed: the $V_\ell$ and $V_b$ distributions become
broader and barely distinguishable, in agreement with the isotropic
model. Please note that some of the broadening in the last bin is
driven by large velocity errors for these distant and faint stars
($G\gtrsim18.8$, see the error-bars in the top-left corner of each
panel). To illustrate the effect of error-blurring on the shape of the
velocity histograms, we deconvolve the distributions using the ED
technique \citep[][]{extreme}. The resulting deconvolved distributions
are shown as dashed lines in Fig.\ \ref{fig:VlVbr}. In the first four
bins, i.e. for $r<30$ kpc, the difference between the observed and
deconvolved distributions is minuscule. In the last two radial bins,
the mismatch is more noticeable but still minor. We conclude that the
change in the shape of the transverse velocity distribution with
radius is dictated primarily by the change in the overall anisotropy
of the stellar halo.

We have checked that the model distributions of the transverse
velocities do not substantially change their behavior as a function of
$r$.  Therefore, we confirm that the radial evolution shown in
Fig.\ \ref{fig:VlVbr} is not compatible with a single kinematic model.
Instead, the data displays a rapid transition from isotropy to radial
anisotropy at small radii around 4 kpc. There appears to be another
transition, at larger distances, beyond 30 kpc, where even though the
velocity errors grow substantially, the RRL motions resemble those
generated by the isotropic model.

{\bf Correlation coefficient evolution.} The change from the orbital
isotropy to the radial anisotropy (and back) is not only apparent in
the amplitudes of the transversal velocities shown in Fig. \ref{fig:VlVbr}
but it is also reflected in the directions of RRL motions displayed
in Fig.\ \ref{fig:Pcoffr}. The Figure reproduces the map shown in the
bottom-right panel of Fig.\ \ref{fig:PMmap} but now as a function of
the Galactocentric distance. The red-blue pattern typical of the
streaming along the radial direction (see the bottom-left panel in
Fig.\ \ref{fig:PMmap}) is already noticeable in the innermost bin,
albeit with a small amplitude. Between 4 and 20 kpc this pattern
becomes much more pronounced (see the bottom-right panel of
Fig.\ \ref{fig:PMmap} for comparison). Outside of 20 kpc, especially
in the last two bins, the correlation coefficient maps lose their
coherence quickly, displaying a mix of random orientations. The
patterns at higher distances should be interpreted with caution, as
the growing velocity errors in these bins will act to destroy
intrinsic correlations, especially the low-amplitude ones.

In conclusion, Fig.\ \ref{fig:VlVbr} and Fig.\ \ref{fig:Pcoffr}
confirm that between 4 and 20-30 kpc from the Galactic centre most of
the stars have proper motions compatible with a model with a strong
radial anisotropy. Outside of this region, the distribution of proper
motions appears to behave similar to the isotropic or, perhaps much
less radially biased, model. This kind of radial trend is expected if
the inner halo is dominated by highly eccentric orbits ($e>0.8$). In
this case, along most of the orbital path, the star's velocity is
predominantly in the radial component.  However, close to the
turn-around points, namely the pericentre and the apocentre, the
radial component of the velocity rapidly decreases to 0. Of course,
the synchronized change in the orbital anisotropy in large portions of
the halo is only possible if the halo is dominated by the tidal debris
from a single accretion event. In this case, the stars naturally
possess similar orbital properties, thus causing the whole halo's
anisotropy to change quickly around their pericentres and apocentres.

Our finding of the switch in the orbital properties of the stellar
halo around 4 kpc is in good agreement with the detection of the
anisotropy change at similar distances presented in \citet{Wegg18} and
with the estimate of the pericenters of the stars belonging to the
Hercules-Aquila and Virgo Clouds in \citet{Simion18}. The decay of the
correlation between the proper motion components beyond 30 kpc is in
agreement with the marked anisotropy decrease measured at similar
distances by \citet{Bird2018} and \citet{Lancaster}. This behavior can
be explained by the traversal of the apocentric pile-up of a large
number of stars sharing a common progenitor as described in
\citet{Deason18}. Moreover, the evolution of the RRL orbital
anisotropy with Galactocentric distance appears to correlate well with
the change in the mixture of RRL periods and amplitudes
\citep[see][]{Unmixing}. Together with these earlier works, out study
gives further support to the hypothesis of \citet{Deason2013} in which
the observed stellar halo density break can be linked to the last
apocentre of a massive progenitor galaxy, whose tidal debris engulfed
the inner Milky Way.

\subsection{Signature of an ancient major merger?}

The simple analysis presented in this Section suggests that the orbits
of the bulk of the RRL stars in the inner halo of the Milky Way
(including the HAC and the VOD) are highly eccentric, in line with the
local measurements of the ``Gaia
Sausage''\citep[see][]{BelokurovSa,MySa,Helmi18}.

 The exact fraction of RRL stars in our sample that can be
  associated with this structure is uncertain.  Recent studies have
  shown that the fraction of the inner halo population with strong
  anisotropy is certainly larger than $50\%$ \citep{Lancaster} and can
  be as high as $70\%$ \citep{Mack18}.  We repeated the analysis
  described in Sec.\ \ref{sec:kine} considering a mixed model with
  both isotropic and strong radially anisotropic velocity
  distributions. We found that the magnitude of the red-blue patterns
  shown by the data in Fig.\ \ref{fig:PMmap} are qualitatively
  reproduced when the fraction of stars belonging to the anisotropic
  population is larger than $70\%$.  Interestingly, several works have
  shown that, during mergers, large numbers of stars in the
  pre-existing disc can be heated to more eccentric and energetic
  orbits forming an in-situ population in the inner halo
  \citep[e.g.][]{Purcell10,Font11}. Although the kinematic properties
  of this kicked-up population may resemble those of the accreted
  debris \citep[see e.g.\ ][]{DiMatteo11,JB17}, the in-situ stars tend
  to maintain a good fraction of their initial (positive) $z$-angular
  momentum \citep[see e.g.\ ][]{Haywood2018}. Since the $z$-angular
  momentum of the \vir{Gaia Sausage} stars is typically very low or
  negative \citep[see][]{BelokurovSa,MySa,Helmi18}, we envisage that
  the contribution of the heated disc stars to our sample is likely
  minimal.
We therefore conclude that the triaxial structure uncovered in
Sec. \ref{sec:slice} is mostly composed of the tidal debris from this
ancient massive merger. While glimpses of the ``Gaia Sausage'' have
recently been found outside of the Solar neighborhood
\citep[see][]{Deason18,Lancaster,Simion18}, ours is the first
comprehensive 3-D map of this gigantic debris structure which
completely dominates the inner stellar halo.

There exists an additional corollary worth pondering about. Depending
on the exact time of accretion and the mass of the progenitor galaxy,
the ``Gaia Sausage'' might also be contributing a significant fraction
of the Dark Matter content in the inner part of the Milky Way. If the
merger delivered some $10^{11} M_{\odot}$ of Dark Matter - which is
not far off most currently available estimates
\citep[see][]{BelokurovSa, Helmi18, Mack18} - then its portion
of the local Dark Matter budget can be gauged as follows. Using the
above progenitor mass and the triaxial model described in the previous
Section, the Dark Matter density at the Solar (elliptical) radius due
to the ``Gaia Sausage'' is $\sim7\times10^{-4}$
M$_{\odot}$pc$^{-3}$. This is approximately $10\%$ of the local Dark
Matter density \citep[see][]{Bovy2012}. Quite probably, however, this
is a lower bound as the density in the central parts of the debris
cloud is over-estimated by our parametric model (see the left-hand
panel of Fig.~\ref{fig:View}). Naturally, we expect a depletion in
density at distances inward of the pericentre. A different estimate
can be obtained by noticing that within 30 kpc, i.e. inside the
apocentre of the ``Gaia Sausage'', the MW mass is estimated to be
$\sim2.5 \times 10^{11}$M$_{\odot}$ \citep[see
  e.g.][]{Gibbons2014,Williams2017,Posti2018}, of which at least $0.5
\times 10^{11}$M$_{\odot}$ can be attributed to the baryonic component
\citep[see][]{McMillan2017}. Therefore, the Dark Matter contribution
associated with the principal component of the stellar halo can be
$\sim50\%$ within 30 kpc.

In this regard, it is curious that the direction of the major axis of
the triaxial stellar halo described in this work is close (within
$20^\circ$) to the direction of the major axis of the triaxial Dark
Matter halo model found by \cite{LM}.  Although the details of the
axial ratios show some difference (the Dark Matter shortest axis is
not on the direction normal to the Galactic plane), the \vir{amount}
of triaxiality as measured by the triaxiality parameter
$T$\footnote{$T=\frac{1-b^2/a^2}{1-c^2/a^2}$, where $a,b,c$ are the
  longest, intermediate and shortest axes, respectively.} is very
similar. While the exact orientation of the Dark Matter halo with
respect to the Galactic disk as suggested by \cite{LM} is not stable,
the triaxial Dark Matter halo and the baryonic disk can co-exist
happily if the disk's spin is allowed to be misaligned with respect to
the principal planes of the potential \citep[see][]{Victor2013}. This
misalignment can be induced, for example, by gas accretion at later
times as demonstrated in \citet{Victor2015}.

\section{Conclusions} \label{sec:conclusions}

In this paper, we take advantage of the most comprehensive sample of
RRL stars to date, i.e. the one supplied by the \gaia DR2, to study
the shape of the Galactic stellar halo. In order to assemble a dataset
with sensible and stable completeness and purity, we had to get rid of
more than a half of the original total of $\sim230,000$ objects. While
this may sound scary, remember that the estimated contamination of the
\gaia DR2 sample shoots up to $\sim 50\%$ in the bulge area
\citep[see][]{GaiaVariable}. For our investigations, we have only kept
RRL stars that possess well-behaved colors and
astrometry. Additionally, we do not consider objects residing in
small-scale over-densities, i.e. known Milky Way satellites and
globular clusters. However, even after all these cuts, we are left
with nearly 93,000 RRL stars across the entire sky, from the inner
several kpc to the distant outskirts - a dataset unprecedented in its
richness and reach.

For the very first time, we produce detailed maps of the RRL density
distribution in the inner Milky Way (see
Fig.~\ref{fig:Zslice}). Slicing the Galaxy at different Galactic
heights, we reveal the evolution of the halo shape from nearly
spherical (close to the Galaxy's centre) to clearly triaxial further
out. Despite several obvious asymmetries and large-scale
over-densities visible in these maps, the overall structure of the
halo appears to be well-described by the SPL-TR$^{qv}$ model of
\citet{Iorio18}. In this model, the major axis lies in the Galactic
plane rotated by $\sim70^{\circ}$ (anti-clockwise) away from the
Galactic $x$-axis, i.e. pointing approximately in the direction of the
Magellanic Clouds. The ratio of the major to intermediate axis is
$p=1.3$, but the ratio of the minor axis (aligned with the Galactic
$z$ axis) to major one changes with distance from the Milky Way centre
(from $q\sim0.5$ to $q\sim0.8$). The clearest deviations from the
above model are the Virgo Over-Density and the Hercules-Aquila Cloud,
which protrude far in the vertical direction on the opposite sides of
the Galactic plane. As we have demonstrated in
Section~\ref{sec:distribution} and Fig.~\ref{fig:View}, the
SPL-TR$^{qv}$ model of \cite{Iorio18} can be adjusted slightly to
encompass both HAC and VOD. This is achieved by bringing its
intermediate axis out of the Galactic plane by $\sim20^{\circ}$. From
the slices presented in Fig.~\ref{fig:Zslice} it is now easy to judge
the whereabouts of the HAC and VOD with respect to the Milky Way's
centre and to each other. These debris clouds appear to be aligned
with the major axis of the triaxial structure discussed above.

The first clues that HAC and VOD may actually be part of the same
accretion event have already been presented by \citet{Simion18}, who
calculated the orbital properties of a sub-sample of RRL stars
residing in these over-densities. Given that both structures appear to
be dominated by stars on extremely eccentric orbits, \citet{Simion18}
conclude that HAC and VOD represent unmixed portions of an ancient
massive head-on collision also known as the ``Gaia Sausage''. In an
attempt to test this hypothesis, we have studied the kinematics of the
inner halo RRL stars as reflected in their proper motions. We claim
that across a wide range of distances from the Sun, the halo's
velocity ellipsoid can be gleaned from its projection on the celestial
sphere. Our results are in good agreement with an earlier analysis of
the RRL proper motion data of \citet{Wegg18}. Compared to the
above study, while we do not attempt to model the shape of the
velocity ellipsoid, we do obtain a broader view of the global
kinematic patterns in the Galactic halo. Accordingly, contrasting the
observed amplitudes and directions of the RRL proper motions
within $\sim30$ kpc from the Galactic centre with simple kinematic
models, we show that the entire inner halo is dominated by stars on
highly eccentric orbits.

We interpret the stretched appearance and the extreme radial
anisotropy of the inner halo as the tell-tale signs of a low angular
momentum collision with a massive satellite
\citep[see e.g.\ ][]{Brook03},
thus associating the bulk
of the RRL stars within 30 kpc with the so-called ``Gaia Sausage''
merger event \citep[][]{BelokurovSa,MySa,Haywood2018,Helmi18}. By
tracking the change in the behavior of the RRL proper motions with
Galactocentric distance, we place constraints on the pericentre
($\lesssim 4$ kpc) and the apocentre (20-40 kpc) of this enormous
tidal debris cloud. The fact that today it appears to be squashed
vertically (in the direction perpendicular to the Galactic disc)
agrees well with the predictions of numerical simulations of the Dark
Matter halo evolution in the presence of baryons
\citep[][]{Kaza2004,Gnedin2004,Debattista2008,Abadi2010}. In terms of
local Dark Matter density, we have estimated the contribution of the
``Gaia Sausage'' to be between $10\%$ and $50\%$. A curious conundrum
is starting to emerge in which a clearly triaxial stellar halo needs
to be reconciled with multiple recent reports of a near spherical
inner Dark Matter halo \citep[][]{Bowden2015,Bovy2016,Wegg18}, a
measurement itself in tension with the results from numerical
simulations which do not produce perfect sphericity
\citep[e.g.][]{Kaza2010,Victor2015}. The spatio-kinematic information
uncovered here will help constrain the total mass as well as the time
of accretion of the ``Gaia Sausage'' progenitor, and thus understand
the role this event played in the life of the Milky Way.

\section*{Acknowledgements}

The authors thank Wyn Evans, Victor Debattista, Andrew Cooper, Sergey
Koposov, Iulia Simion, Alis Deason for most illuminating discussions
which helped to improve this manuscript. G.I. is supported by the
Royal Society Newton International Fellowship.  The research leading
to these results has received funding from the European Research
Council under the European Union's Seventh Framework Programme
(FP/2007-2013) / ERC Grant Agreement n. 308024.  This work has made
use of data from the European Space Agency (ESA) mission {\it Gaia}
(\url{https://www.cosmos.esa.int/gaia}), processed by the {\it Gaia}
Data Processing and Analysis Consortium (DPAC,
\url{https://www.cosmos.esa.int/web/gaia/dpac/consortium}). Funding
for the DPAC has been provided by national institutions, in particular
the institutions participating in the {\it Gaia} Multilateral
Agreement.


\bibliographystyle{mnras}
\bibliography{references}



\bsp	
\label{lastpage}
\end{document}